\documentclass[12pt]{article}
\usepackage{graphicx}
\usepackage{cite}

\def \bea{\begin{eqnarray}}
\def \beq{\begin{equation}}
\def \b{{\cal B}}

\def \eea{\end{eqnarray}}
\def \eeq{\end{equation}}

\def \s{\sqrt{2}}

\def \half{\frac{1}{2}}

\def \3half{\frac{3}{2}}

\begin{document}
\begin{flushright}
EFI 08-13 \\
May 2008 \\
arXiv:0805.4601 \\
\end{flushright}
\centerline{\bf Small amplitude effects  in $B^0\to D^+D^-$ and related decays}
\bigskip
\centerline{Michael Gronau\footnote{On sabbatical leave from the Physics 
Department, Technion, Haifa 32000, Israel.}, Jonathan L. Rosner}
\medskip
\centerline{\it Enrico Fermi Institute and Department of Physics,
 University of Chicago} 
\centerline{\it Chicago, IL 60637, U.S.A.} 
\medskip
\centerline{Dan Pirjol}
\centerline{\it National Institute for Physics and Nuclear Engineering, Department of Particle Physics}
\centerline{\it 077125 Bucharest, Romania}
\bigskip
\begin{quote}
Intrigued by a recent Belle result for a large direct CP asymmetry in $B^0\to
D^+D^-$, we study the effects of a $\bar b\to \bar uu\bar d$ quark transition  
by combining the asymmetry information with rates and asymmetries in
isospin-related decays. 
Arguing for a hierarchy among several contributions to these decays, including
an exchange amplitude which we estimate, we present tests for factorization of
the leading terms, and obtain an upper bound on the ratio 
of $\bar b\to \bar uu\bar d$ and $\bar b\to \bar cc\bar d$ amplitudes. 
We prove an approximate $\Delta I=1/2$ amplitude relation for $B\to D\bar D$,
and an approximate equality between CP asymmetries in $B^0\to D^+D^-$ and 
$B^+\to D^+\bar D^0$.  Violations of these relations by Belle measurements, at
$1.8\sigma$ and $3.6\sigma$ respectively, if confirmed, would
indicate a possible New Physics
contribution in $\bar b\to \bar uu\bar d$.  Applying flavor SU(3), we extend
this study to a total of ten processes including $\Delta S=0$ decays involving
final $D_s$ and initial $B_s$ mesons, and $\Delta S=1$ decays of $B$ and $B_s$
mesons into pairs of charmed pseudoscalar mesons.  The decays $B_s\to D\bar D$
provide useful information about a small exchange amplitude, responsible for a
decay rate difference between $B^+\to D^+\bar D^0$ and $B^0\to D^+D^-$.
A method for determining the weak phase $\gamma$, based on CP asymmetries in
$B^0(t)\to D^+D^-$ and the decay rate for $B_s\to D^+_sD^-_s$ or $B^{+(0)}\to
D^+_s\bar D^0 (D^-)$, is shown to involve high sensitivity to SU(3) breaking. 
\end{quote} 

\leftline{\qquad PACS codes:  12.15.Hh, 12.15.Ji, 13.25.Hw, 14.40.Nd} 

\bigskip
\section{Introduction}
Accurate measurements of the weak phase $\beta\equiv {\rm arg}(V^*_{tb}V_{td}
/V^*_{cb}V_{cd})$, $\sin 2\beta= 0.680\pm 0.025,~\cos
2\beta>0$~\cite{Barberio:2007cr}, have provided a precision test for the
Cabibbo-Kobayashi-Maskawa \cite{Cabibbo:1963yz,Kobayashi:1973fv} framework 
and for the Kobayashi-Maskawa mechansim of CP violation.  The accuracy of this
test relies on the pure dominance by a single weak phase of  a few $\bar b\to
\bar cc\bar s$ processes including $B^0\to
J/\psi K_S$~\cite{Gronau:1989ia,Boos:2004xp}.  This implies a mixing-induced
asymmetry, $S=\sin 2\beta$, and a vanishingly small direct CP asymmetry, 
as confirmed experimentally, $A_{CP}=-0.012\pm 0.020$~~\cite{Barberio:2007cr}.

The decay $B^0\to D^+ D^-$ is dominated by $\bar b\to \bar cc\bar d$, but
involves a smaller non-negligible amplitude from $\bar b\to\bar uu\bar d$
carrying a different weak phase.  The second amplitude introduces hadronic
uncertainties in predictions for the asymmetries $S$ and $A_{CP}$ in this
process.  Early model-independent estimates of the ratio of the two amplitudes
contributing to this process  vary from a few percent to upper bounds of about
$0.2$~\cite{Gronau:1989ia,Grinstein:1989df} or  $0.3$~\cite{Ciuchini:1997zp}.
A more recent model-dependent calculation finds 0.03~\cite{Xing:1999yx}. Values
larger than 0.3 may be obtained in extensions of the Standard
Model~\cite{Grossman:1996ke}. The two asymmetries depend also on the strong
phase difference between the  $\bar b\to \bar cc\bar d$ and $\bar b\to\bar
uu\bar d$ amplitudes~\cite{Gronau:1989ia}. 

\begin{table}
\begin{center}
{\begin{tabular}{ccccc} 
\hline\hline
Decay mode&~&BaBar& Belle&Average\\
\hline
$B^0 \to D^+D^-$&$\b$&$2.8\pm 0.4\pm 0.5$&$1.97\pm 0.20\pm 0.20$
&$2.11\pm 0.31$\\
~&$A_{CP}$&$-0.11\pm 0.22\pm 0.07$&$0.91\pm 0.23\pm 0.06$
&$0.37\pm 0.17$\\
~&$S$& $-0.54\pm 0.34\pm 0.06$&$-1.13\pm 0.37\pm 0.09$
&$-0.75\pm 0.26$\\
\hline
$B^+\to D^+\bar D^0$&$\b$&$3.8\pm 0.6\pm 0.5$&$3.85\pm 0.31\pm 0.38$
&$3.84\pm 0.42$\\
&$A_{CP}$&$-0.13\pm 0.14\pm 0.02$&$0.00\pm 0.08\pm 0.02$
&$-0.03\pm 0.07$\\
\hline
$B^0\to D^0\bar D^0$&$\b$&$<0.6~(90\%$~c.~l.)&$<0.43~(90\%$~c.~l.)
&$<0.43~(90\%$~c.~l.)\\
\hline
$B^0\to D^+_sD^-_s$&$\b$&$<1.0~(90\%$~c.~l.)&  $< 0.36~(90\%$~~c.~l.)
 &$<0.36~(90\%$~c.~l.)\\
\hline\hline
\end{tabular}}
\caption{Charge-averaged branching ratios $\b$ in units of $10^{-4}$ 
and CP asymmetries $A_{CP}, S$ in $B\to D\bar D$, from
Refs.~\cite{Aubert:2006ia,Fratina:2007zk,Aubert:2007pa,Adachi}.
Also included are upper limits on 
$\b(B^0\to D^+_sD^-_s)$~\cite{Aubert:2005jv,Zupanc:2007pu}.}
\label{tab:1}
\end{center}
\end{table}

Asymmetries in $B^0\to D^+D^-$, measured by the BaBar and Belle collaborations,
are quoted in the upper part of Table I.  The table also includes branching
ratios for $B^0\to D^+D^-, B^+\to D^+\bar D^0, B^0\to D^0\bar D^0$, a direct CP
asymmetry measured for $B^+\to D^+\bar D^0$~\cite{Aubert:2006ia,%
Fratina:2007zk,Aubert:2007pa,Adachi}, and upper limits on
$\b(B^0\to D^+_sD^-_s)$ measured by BaBar~\cite{Aubert:2005jv}
and Belle~\cite{Zupanc:2007pu}.  The BaBar asymmetries in
$B^0\to D^+D^-$~\cite{Aubert:2007pa} are consistent with $A_{CP}=0, S=-\sin
2\beta$, showing no evidence for a $\bar b\to \bar uu\bar d$ term in the decay
amplitude. In contrast, the Belle asymmetry 
measurements~\cite{Fratina:2007zk}, which fluctuate outside the physical
region, $A^2_{CP}+S^2\le 1$,  deviate substantially from the above nominal
values, indicating a sizable $\bar b\to\bar uu\bar d$ amplitude.  The direct
asymmetry $A_{CP}$ measured by Belle  is nonzero at a level higher than
$3\sigma$. Its central value indicates the possibility of a second amplitude
larger than permitted in the Cabibbo-Kobayashi-Maskawa (CKM) framework.

A major goal of this paper, largely intrigued by the Belle results, is to study carefully
the dynamics and CKM structure of the $B^0\to D^+D^-$ decay amplitude and of
decay amplitudes for the two isospin-related processes, $B^+\to D^+\bar D^0$
and  $B^0\to D^0\bar D^0$.  In references~\cite{Xing:1999yx} and
\cite{Sanda:1996pm} these processes have been stated to originate in a $\Delta
I=1/2$ effective Hamiltonian implying an isospin triangle relation among the
three amplitudes.  It will be shown that, while $\Delta I=1/2$ is not a
property of the effective Hamiltonian, an approximate $\Delta I=1/2$ rule is
expected to hold for the three decay amplitudes and should be tested
experimentally.  Applying flavor SU(3) to the above processes, we will extend
our study to include strangeness-conserving decays involving final $D_s$ and
initial $B_s$ mesons, and strangeness-changing decays of $B$ and $B_s$ mesons
into pairs of charmed pseudoscalar mesons.

In Section 2 we study the asymmetries measured by BaBar and Belle in 
$B^0\to D^+D^-$ in terms of two parameters, the ratio $r$ of   $\bar b\to \bar uu\bar d$ and 
$\bar b\to \bar cc\bar d$ amplitudes and their relative strong phase
$\delta$. Section 3 introduces 
expressions for the amplitudes of the three processes $B^0\to D^+D^-, B^0\to D^0\bar D^0$ 
and $B^+\to D^+\bar D^0$ in terms of isospin amplitudes.
We identify circumstances under which an isospin triangle relation 
between these amplitudes can be violated by a (small) $\Delta I=3/2$ contribution.
Section 4 studies $B\to D\bar D$ decays and two other SU(3) related $\Delta S=0$ decays
of $B^0$ and $B_s$ in terms of graphical contributions, while Section 5 
extends this study to corresponding strangeness changing decays  of $B^0, B^+$ and $B_s$. 
Section 6 discusses a hierarchy among graphical amplitudes, presenting tests of factorization
for the dominant terms. In Section 7 we discuss briefly consequences of this hierarchy on a
theoretical upper limit on $r$, illuminating an inconsistency between the CP asymmetries
measured by Belle in $B^0\to D^+D^-$ and $B^+\to D^+\bar D^0$.
Section 8 discusses a way for determining the weak phase $\gamma$ by combining 
information from asymmetries in $B^0\to D^+D^-$ and decay rates of corresponding 
$\Delta S=1$ decays to charm-anticharm, while Section 9 concludes. An Appendix 
provides a dictionary between graphical amplitudes and SU(3) reduced matrix elements 
of four-quark operators appearing in the effective Hamiltonian.

\section{Ratio of $\bar b\to \bar uu\bar d$ and $\bar b\to \bar cc\bar d$ terms 
in $B^0\to D^+D^-$}\label{r-delta}

We start our discussion by translating the  $B^0\to D^+D^-$ asymmetries, 
measured separately by BaBar and Belle, into values of the ratio $r$ of 
$\bar b\to\bar uu\bar d$ and $\bar b\to \bar cc\bar d$ amplitudes and the 
relative strong phase $\delta$ between these amplitudes. 
Denoting 
\bea
A(B^0\to D^+D^-) & = & A_c + A_u\,e^{i(\delta + \gamma)} = 
A_c\left[ 1 + r\,e^{i(\delta + \gamma)}\right]~,~~~~~(r\equiv A_u/A_c)~,
\nonumber\\
A(\bar B^0 \to D^+D^-) & = & A_c + A_u\,e^{i(\delta - \gamma)} = 
A_c\left[ 1 + r\,e^{i(\delta - \gamma)}\right]~,
\nonumber\\
\lambda_{D^+D^-} & \equiv & e^{-2i\beta}\frac{A(\bar B^0 \to D^+D^-)}{A(B^0\to D^+D^-) }~,
\label{Amp}
\eea
one has~\cite{Gronau:1989ia}
\bea
S(D^+D^-) &\equiv& \frac{2{\rm Im}(\lambda_{D^+D^-})}{1+|\lambda_{D^+D^-}|^2}
= - \frac{\sin 2\beta +2r\cos\delta\sin(2\beta+\gamma)+r^2\sin 2(\beta+\gamma)}
{1+2r\cos\delta\cos\gamma +r^2}~,
\nonumber\\
A_{CP}(D^+D^-) &\equiv& \frac{|\lambda_{D^+D^-}|^2-1}{|\lambda_{D^+D^-}|^2+1}
= \frac{2r\sin\delta\sin\gamma}{1+2r\cos\delta\cos\gamma +r^2}~.
\label{SACP-exact}
\eea
Keeping only linear terms in $r$, 
\bea
S(D^+D^-) & \simeq & -\sin 2\beta - 2r\cos 2\beta\cos\delta\sin\gamma~, 
\nonumber\\
A_{CP}(D^+D^-) & \simeq & 2r\sin\delta\sin\gamma~,
\label{SACP-approx}
\eea
implies
\beq
\label{r-approx}
r\approx \frac{\sqrt{[(S+\sin 2\beta)/\cos 2\beta]^2 + A^2_{CP}}}{2\sin\gamma}~.
\eeq

Consider the measured asymmetries and the current values of  $\beta$ and $\gamma$,  
$\beta = (21.5\pm 1.0)^\circ$~\cite{Barberio:2007cr},  
$\gamma=(67.6^{+2.8}_{-4.5})^\circ$~\cite{CKMfitter} (see also~\cite{UTfit,Gronau:2007xg}).
Using this information, the approximation (\ref{SACP-approx}) and 
(\ref{r-approx}), or the precise expressions (\ref{SACP-exact}), determine $r$ and $\delta$. 
The resulting errors in $r$ and $\delta$ are dominated by the errors in the measured asymmetries. Taking central values for the asymmetries and values $\beta =21.5^\circ, \gamma = 68^\circ$, 
Eq.~(\ref{r-approx}) implies central values around $r=0.1$ (BaBar) and $r=0.6$ (Belle). 
In both cases the error in $r$  is about $0.2$. The central value  of $r$ for Belle, for which the linear approximations (\ref{SACP-approx}) involve non-negligible quadratic corrections, should be 
considered with care because this value of $r$ is based on non-physical values of the asymmetries obeying $A^2_{CP}+S^2>1$.  

\begin{figure}[t!]
\centering{\includegraphics[width=8.0cm]{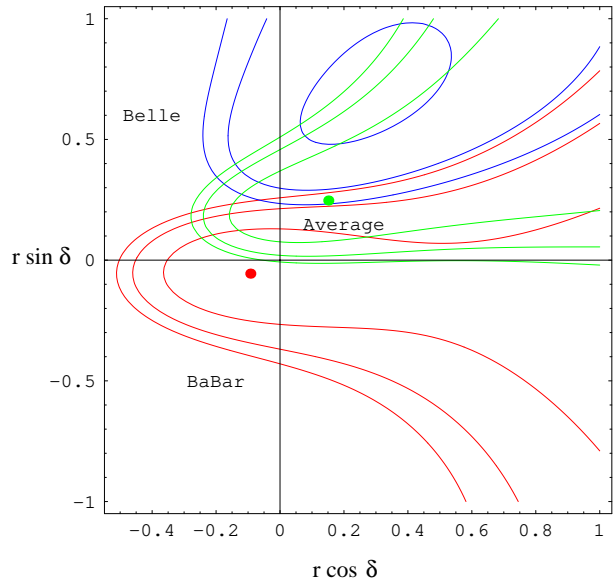}}
\caption{$\chi^2$ plots in the $(r\cos\delta,r\sin\delta)$ plane assuming $\beta =21.5^\circ, \gamma=68^\circ$. The red, blue and green curves show constraints following
from $B^0\to D^+D^-$ asymmetries measured by BaBar, Belle and  their averages. 
In each case the most inside,  intermediate and most outside curves represent bounds at
$68\%, 90\%$ and $95\%$ confidence levels. Red and green points describe solutions 
corresponding to central values of the BaBar asymmetries and the averaged asymmetries.}
\label{fig:rdelta-chi2}
\end{figure}

In order to study the implications of the asymmetries on the pair of parameters  ($r, \delta$), 
we have performed a two dimensional $\chi^2$ analysis for these two parameters using 
the asymmetry measurements and assuming $\beta =21.5^\circ, \gamma=68^\circ$. In Fig.~\ref{fig:rdelta-chi2} we plot the resulting contours in the $(r\cos\delta,r\sin\delta)$ 
plane for $\chi^2=2.30, 4.61, 5.99$, corresponding to $68\%, 90\%, 95\%$
confidence levels. The red, blue and green curves show constraints following
from the asymmetries measured by BaBar, Belle and  their averages. In each case the 
innermost,  intermediate and outermost curves describe bounds at $68\%, 90\%$ 
and $95\%$  confidence levels. The red and green points are solutions corresponding
to central values of the BaBar asymmetries and the averaged asymmetries. We do not
show a central Belle point because the central values of the Belle asymmetries lie outside the 
physical region.  

The Belle  two-parameter boundary curve for $90\%$ confidence level contains a point 
with closest distance to the origin, $r=0.29, \delta=82^\circ$. The point on this curve with 
smallest $\delta$ has $\delta=37^\circ, r=0.87$. (Note that the values $r=0.29$ and 
$\delta=37^\circ$ are lower than the $90\%$ confidence level lower limits on these 
separate single variables.)
Thus, the Belle data alone would provide evidence 
for a sizable $r$ and for a large strong phase difference $\delta$. 
We note, however, that the BaBar and Belle regions of $90\%$ confidence level do not 
overlap. In order to draw firm conclusions about $r$ and $\delta$ one should therefore 
wait for better agreement between the asymmetries measured by the two collaborations.

\section{Isospin amplitudes in $B\to D\bar D$}\label{isospin}

The low energy effective Hamiltonian governing $B^0\to D^+D^-$, 
$B^0\to D^0\bar D^0$ and $B^+\to D^+\bar D^0$ involves two CKM factors $V^*_{cb}V_{cd}$ 
and $V^*_{ub}V_{ud}$, both of order $\lambda^3$ 
($\lambda=|V_{us}|=0.2258\pm 0.0010$~\cite{CKMfitter}). 
Each of these factors multiplies a combination of four quark operators with coefficients given by calculable Wilson coefficients $C_i$~\cite{Buchalla:1995vs},  
\beq\label{Heff}
{\cal H}_{\rm eff} = \frac{G_F}{\s}\left[V^*_{cb}V_{cd}\sum_{i=1}^2C_i{\cal O}^c_i +
V^*_{ub}V_{ud}\sum_{i=1}^2C_i{\cal O}^u_i + (V^*_{cb}V_{cd} + V^*_{ub}V_{ud})
\sum_{k=3}^{10}C_k{\cal O}_k\,\right]~.
\eeq
The current-current operators ${\cal O}_i^c$ and ${\cal O}_i^u$ ($i=1,2$) have flavor dependence 
$(\bar b c)(\bar c d)$ and $(\bar b u)(\bar u d)$, respectively. 
Thus, while the first  pair of operators are pure $\Delta I=\half$, the second pair involves 
both $\Delta I=\half$ and $\Delta I=\3half$. The QCD penguin operators ${\cal O}_k$ ($k=3-6$)
with flavor structure $(\bar b d)$ are pure $\Delta I=1/2$, while the electroweak 
penguin operators ${\cal O}_k\sim (\bar b d)\sum_qe_q(\bar q q)$ ($k=7-10,~q=u,d,s,c$), 
which depend on the quark charges $e_q$, involve both 
$\Delta I=1/2$ and $\Delta I=3/2$. Thus, in contrast to statements made in 
Refs.~\cite{Xing:1999yx} and~\cite{Sanda:1996pm}, the effective Hamiltonian underlying 
$B\to D\bar D$ decays is not pure $\Delta I=1/2$. It contains an additional $\Delta I=3/2$ component,   
also when neglecting electroweak penguin contributions which have very small Wilson 
coefficients~~\cite{Buchalla:1995vs}.

The final $D\bar D$ states  consist of $I=0$ and $I=1$. This implies a total of  three isospin 
amplitudes, $A_{0,\half}, A_{1,\half}$ and $A_{1,\3half}$, where the two subscripts denote 
$I(D\bar D)$ and $\Delta I$, respectively. Neglecting very small electroweak penguin contributions, 
the $\Delta I=3/2$ amplitude $A_{1,\3half}$ occurs in association with a CKM factor $V^*_{ub}V_{ud}$
but not with $V^*_{cb}V_{cd}$.

The three physical $B\to D\bar D$ decay amplitudes can be written in terms of the 
three isospin amplitudes,
\bea
A^{+-} & \equiv & A(B^0\to D^+D^-) =  \half\,A_{0,\half} + \half\,A_{1,\half} + \half\,A_{1,\3half}~,
\nonumber\\
A^{00} & \equiv & A(B^0\to D^0\bar D^0) =  -\half\,A_{0,\half} + \half\,A_{1,\half} + \half\,A_{1,\3half}~,
\nonumber\\
A^{+0} & \equiv & A(B^+\to D^+\bar D^0) =  A_{1,\half} - \half\,A_{1,\3half}~.
\eea
These relations can be inverted,
\bea
A_{0,\half} &=& A^{+-} - A^{00}~,
\nonumber\\
A_{1,\half} &=& \frac{1}{3}(A^{+-} + A^{00} + 2A^{+0})~,
\nonumber\\
A_{1,\3half} &=& \frac{2}{3}(A^{+-} + A^{00} - A^{+0})~.
\eea

While $\Delta I=1/2$ is not a property of the low energy effective Hamiltonian, we will 
argue below that  $|A_{1,\3half}|\ll |A_{0,\half}|, |A_{1,\half}|$ is a reasonable 
assumption which should be tested experimentally.  
Neglecting the $\Delta I=3/2$ amplitude,  one obtains an approximate 
triangle relation~\cite{Xing:1999yx,Sanda:1996pm},
\beq\label{triangle}
A^{+-} + A^{00} = A^{+0}~.
\eeq
A potential proof for a nonzero $\Delta I=3/2$ amplitude would be a violation of (\ref{triangle}).
This happens when the amplitude triangle does not close, for instance when 
$|A^{+-}| + |A^{00}| < |A^{+0}|$.

In order to illustrate such a possibility consider the Belle measurements for 
$B\to D\bar D$. We define
\bea
|A^{+-}| &=& 10^2\,\sqrt{\b(B^0\to D^+D^-)\left[1-A_{CP}(B^0\to D^+D^-)\right]}~,
\nonumber\\
|A^{+0}| &=& 10^2\,\sqrt{\b(B^+\to D^+\bar D^0)\left[1-A_{CP}(B^+\to D^+\bar D^0)\right]
(\tau_0/\tau_+)}~,
\nonumber\\
|A^{00}| &=& 10^2\,\sqrt{\b(B^0\to D^0\bar D^0)\left[1-A_{CP}(B^0\to D^0\bar D^0)\right]}~.
\eea
Using the Belle values for CP-averaged branching ratios and CP asymmetries given in 
Table  \ref{tab:1}, and a ratio of $B^+$ and $B^0$ lifetimes~\cite{Barberio:2007cr} 
$\tau_+/\tau_0 = 1.071\pm 0.009$, we find
\beq\label{num-triangle}  
|A^{+-}| = 0.42 \pm 0.56~,~~~|A^{+0}| = 1.90 \pm 0.14~,~~~|A^{00}|< 0.57~(1\sigma)~.
\eeq

The small magnitude of $A^{+-}$ follows from the large positive CP asymmetry
measured by Belle, implying observing mostly $\bar B^0$ decays into $D^+D^-$
with only a few $B^0$ decays into this final state. 
The upper bound on $|A^{00}|$ is obtained from a $1\sigma$ upper limit on
$\b(B^0\to D^0\bar D^0)$.  We have assumed that the CP asymmetry in
$B^0\to D^0\bar D^0$ is not large and negative
(in Section 7 we will argue for a vanishingly small asymmetry), 
and we have neglected possible correlations between 
errors in branching ratio and asymmetry measurements in the other two modes. 
We note that the triangle (\ref{triangle}) does not close for central values
of the measured amplitudes (\ref{num-triangle}), and allowing 
for deviations from these values up to $1.8\sigma$ . 

In contrast, the triangle relation holds well when using the central values of
the BaBar measurements. A critical test for the closure of the amplitude
triangle requires a reduction in errors and a better agreement between BaBar
and Belle.

\section{$\Delta S=0$ decays into $D\bar D$ using graphical amplitudes}
\label{DS=0}

Useful expressions for amplitudes in $B \to D\bar D$ using flavor SU(3), which can 
be generalized to $B^0 \to D^+_sD^-_s$ and $B_s \to D^+D^-_s$,  are in terms of 
graphical contributions representing the flow of isospin and flavor SU(3) in these decays~\cite{Gronau:1994rj}. This includes a (color-favored) tree amplitude $T$ 
involving $V^*_{cb}V_{cd}$, penguin and penguin-annihilation
amplitudes $P$ and $PA$ with $u,c$ and $t$-quark loops, exchange amplitudes $E_c$ and 
$E_u$ involving $V^*_{cb}V_{cd}$ and $V^*_{ub}V_{ud}$, and an annihilation amplitude 
$A_{u}$ involving $V^*_{ub}V_{ud}$. The amplitude $E_c$ is associated with popping a 
pair of $u\bar u$ or $d\bar d$ out of the vacuum, 
while $E_u$ and $A_u$ involve $c\bar c$ popping. At this point we neglect very small 
electroweak penguin contributions to which we return in the next section. 

The graphical amplitudes have well-defined isospin properties.
The two graphical amplitudes $E_u$ and $A_u$ involve both $\Delta I=1/2$ and $\Delta I=3/2$, 
while all other amplitudes are pure $\Delta I=1/2$ by construction. 
We will show below that  the $\Delta I=3/2$ amplitude consists solely of the combination 
$E_u +A_u$. This combination is thus responsible for a potential violation of the amplitude 
triangle relation (\ref{triangle}).

We denote
\beq
P = V^*_{ub}V_{ud}\,p_u + V^*_{cb}V_{cd}\,p_c + V^*_{tb}V_{td}p_t= 
V^*_{cb}V_{cd}\,p_{ct} + V^*_{ub}V_{ud}\,p_{ut}~,~~~(p_{ij}\equiv p_i-p_j)~,
\eeq
absorbing  the first term in the tree amplitude by defining,
\beq\label{t_c}
V^*_{cb}V_{cd}\,t_c \equiv T + V^*_{cb}V_{cd}\,p_{ct}~.
\eeq
Similarly, writing
\beq
PA =  V^*_{ub}V_{ud}\,pa_u + V^*_{cb}V_{cd}\,pa_c + V^*_{tb}V_{td}pa_t=
V^*_{cb}V_{cd}\,pa_{ct} + V^*_{ub}V_{ud}\,pa_{ut}~,~~~(pa_{ij}\equiv pa_i -pa_j)~,
\eeq
the first term is absorbed in $E_c$,
\beq\label{e_c}
V^*_{cb}V_{cd}\,e_c \equiv E_c + V^*_{cb}V_{cd}\,pa_{ct}~.
\eeq
Other terms involving the CKM factor $V^*_{ub}V_{ud}$ are 
\beq
E_u \equiv V^*_{ub}V_{ud}\,e_u~, ~~~A_u \equiv V^*_{ub}V_{ud}\,a_u~.
\eeq

Using these definitions with  a shorthand notation, $p_u\equiv p_{ut}, pa_u\equiv pa_{ut}$, we find:
\bea\label{5amps}
a.~~~~~A(B^0 \to D^+D^-) &=& V^*_{cb}V_{cd}\,(t_c + e_c)  + V^*_{ub}V_{ud}\,(p_u + pa_u)~,
\nonumber\\
b.~~~~~~A(B^0 \to D^0 \bar D^0) &=& V^*_{cb}V_{cd}\,(-e_c)  + V^*_{ub}V_{ud}\,(-pa_u - e_u)~,
\nonumber\\
c.~~~~~A(B^+ \to D^+\bar D^0) &=& V^*_{cb}V_{cd}\,(t_c)  + V^*_{ub}V_{ud}\,(p_u + a_u)~,
\nonumber\\
d.~~~~~A(B^0 \to D^+_s D^-_s) &=& V^*_{cb}V_{cd}\,(e_c)  + V^*_{ub}V_{ud}\,(pa_u)~,
\nonumber\\
e.~~~~~A(B_s \to D^+D^-_s) &=& V^*_{cb}V_{cd}\,(t_c)  + V^*_{ub}V_{ud}\,(p_u)~.
\eea
The minus sign in the amplitude involving a $D^0$ follows from our convention,
$D^0\equiv -c\bar u$~\cite{Gronau:1994rj}.

The five physical amplitudes depend on the four combinations, $V^*_{cb}V_{cd}\,t_c$ 
$+~V^*_{ub}V_{ud}\,p_u$, $V^*_{cb}V_{cd}\,e_c  + V^*_{ub}V_{ud}\,pa_u, V^*_{ub}V_{ud}\,e_u,
V^*_{ub}V_{ud}\,a_u$. This equals the number of independent SU(3) reduced matrix elements 
describing $\Delta S=0$ $B$ and $B_s$ decays to pairs of charmed pseudoscalar mesons,
$\langle 1|\!|\overline{\mathbf 3}|\!|3\rangle, \langle 8|\!|\overline{\mathbf 3}|\!|3\rangle, 
\langle 8|\!|\mathbf{6}|\!|3\rangle, \langle 8|\!|\overline{\mathbf 15}|\!|3\rangle$ (see Appendix A). Consequently the five decay 
amplitudes are not mutually independent. They obey one triangle relation in the SU(3) 
symmetry limit,
\beq\label{SU3tri}
A(B^0 \to D^+_s D^-_s) + A(B_s \to D^+D^-_s) = A(B^0 \to D^+D^-)~.
\eeq

The parameters $r$ and $\delta$ studied in Section~\ref{r-delta} can be expressed in terms 
of graphical amplitudes,
\beq\label{r,del}
r = \frac{|V^*_{ub}V_{ud}|}{|V^*_{cb}V_{cd}|}\frac{|p_u + pa_u|}{|t_c + e_c|}~,~~~~~~
\delta = {\rm arg}\left(-\frac{p_u + pa_u}{t_c + e_c}\right)~,
\eeq
and the isospin amplitudes defined in section \ref{isospin} are given by
\bea
A_{0,\half} &=& V^*_{cb}V_{cd}\,(t_c + 2e_c) + V^*_{ub}V_{ud}\,(p_u + 2pa_u +e_u)~,
\nonumber\\
A_{1,\half} &=& V^*_{cb}V_{cd}\,t_c  + \frac{1}{3}V^*_{ub}V_{ud}\,(3p_u -e_u + 2a_u)~,
\nonumber\\
\label{3half}
A_{1,\3half} &=& -\frac{2}{3}V^*_{ub}V_{ud}\,(e_u + a_u)~.
\eea
The last relation confirms our statement above  that the $\Delta I=3/2$ amplitude involves 
only the combination $E_u + A_u$.

\section{$\Delta S=1$ $B$ and $B_s$ decays into charm-anticharm}\label{DS=1}

The parametrization (\ref{5amps}) of $\Delta S=0$ decays into charm-anticharm in terms of 
flavor SU(3) graphical amplitudes can be extended to $\Delta S=1$ CKM-favored decays of 
$B$ and $B_s$ mesons, which are governed by $\bar b \to \bar cc\bar s$. U-spin reflection 
symmetry $d\leftrightarrow s$ implies expressions similar to Eqs.~(\ref{5amps}) for 
corresponding U-spin related amplitudes~\cite{Gronau:2000zy}, in which one replaces $V^*_{cb}V_{cd}$ by $V^*_{cb}V_{cs}$ and $V^*_{ub}V_{ud}$ by $V^*_{ub}V_{us}$. 
Since U-spin transforms $B^0\leftrightarrow B_s,~D^{\pm}\leftrightarrow D^{\pm}_s$ while 
keeping $B^+, D^0$ and $\bar D^0$ invariant, one has:
\bea\label{5amps'}
a.~~~~~A(B_s \to D^+_sD^-_s) &=& V^*_{cb}V_{cs}\,(t_c + e_c)  + V^*_{ub}V_{us}\,(p_u + pa_u)~,
\nonumber\\
b.~~~~~~A(B_s \to D^0 \bar D^0) &=& V^*_{cb}V_{cs}\,(-e_c)  + V^*_{ub}V_{us}\,(-pa_u - e_u)~,
\nonumber\\
c.~~~~~A(B^+ \to D^+_s\bar D^0) &=& V^*_{cb}V_{cs}\,(t_c)  + V^*_{ub}V_{us}\,(p_u + a_u)~,
\nonumber\\
d.~~~~~A(B_s \to D^+ D^-) &=& V^*_{cb}V_{cs}\,(e_c)  + V^*_{ub}V_{us}\,(pa_u)~,
\nonumber\\
e.~~~~~A(B^0 \to D^+_sD^-) &=& V^*_{cb}V_{cs}\,(t_c)  + V^*_{ub}V_{us}\,(p_u)~.
\eea

We note that while both $V^*_{cb}V_{cd}$ and $V^*_{ub}V_{ud}$ in (\ref{5amps}) are 
of order $\lambda^3$, the factors $V^*_{cb}V_{cs}$ and $V^*_{ub}V_{us}$ in (\ref{5amps'}) 
are order $\lambda^2$ and $\lambda^4$ respectively. More precisely, one has
\beq
-\frac{V^*_{cb}V_{cd}}{V^*_{cb}V_{cs}} = \frac{V^*_{ub}V_{us}}{V^*_{ub}V_{ud}}=
\frac{\lambda}{1-\lambda^2/2}~.
\eeq
This implies that in the U-spin symmetry limit CP rate differences in the five U-spin pairs of corresponding $\Delta S=0$ and $\Delta S=1$ decays, such as $B^0 \to D^+D^-$ and 
$B_s \to D^+_sD^-_s$, have equal magnitudes but opposite signs~\cite{Gronau:2000zy},
\beq\label{BBs}
\Gamma(\bar B_s \to D^+_sD^-_s) - \Gamma(B_s \to D^+_sD^-_s) =
-[\Gamma(\bar B^0 \to D^+D^-) - \Gamma(B^0 \to D^+D^-)]~.
\eeq
Since the rates of $\Delta S=1$ decays  are about $\lambda^{-2}$ larger than those 
of $\Delta S=0$ decays, the CP asymmetries in the former are expected to be about 
$\lambda^2$ smaller than in the latter. The case of Eq.~(\ref{BBs}) has been discussed in 
Ref.~\cite{Fleischer:1999nz}. Similar CP rate difference relations hold between the
other four pairs of processes.
An SU(3) triangle relation analogous to (\ref{SU3tri}) holds for $\Delta S=1$ decays,
\beq
A(B_s \to D^+ D^-) + A(B^0 \to D^+_sD^-) = A(B_s \to D^+_sD^-_s)~.
\eeq
  
So far we have neglected electroweak penguin contributions in both $\Delta S=0$
and $\Delta S=1$ decays. This is justifiable in the first case by very small Wilson 
coefficients associated with electroweak penguin operators~\cite{Buchalla:1995vs}
multiplying a CKM factor $V^*_{tb}V_{td}$ of the usual order $\lambda^3$. However, 
electroweak penguin terms in strangeness-changing decays involve a CKM factor 
$V^*_{tb}V_{ts}\sim {\cal O}(\lambda^2)$, much larger than terms involving
$V^*_{ub}V_{us}\sim {\cal O}(\lambda^4)$ which are kept in (\ref{5amps'}). 
Thus, for consistency one must keep also electroweak penguin terms in $\Delta S=1$ 
decays. 

In general, there are four types of diagrams describing electroweak penguin (EWP) contributions,
corresponding to the above-mentioned four SU(3) reduced matrix elements.
The four diagrams may be associated with a color-suppressed EWP amplitude $P^C_{EW}$,
an EWP-exchange amplitude $PE_{EW}(c)$ associated with $c\bar c$ popping,  and 
two EWP-annihilation amplitudes, $PA_{EW}$ and $PA_{EW}(c)$, associated with 
$u\bar u, d\bar d, s\bar s$ popping and $c\bar c$ popping, respectively.
We neglect the two EWP amplitudes involving $c\bar c$ popping for a reason discussed in Section \ref{hierarchy}. Using $V^*_{tb}V_{td(s)}= - V^*_{cb}V_{cd(s)}-V^*_{ub}V_{ud(s)}$, the remaining 
two EWP amplitudes, $P_{EW}^C$ and $PA_{EW}$, may be 
absorbed in the following way in definitions of four amplitudes occurring in (\ref{5amps}) and (\ref{5amps'}) without changing these ten equations:
\beq\label{EWP-trans}
t_c - \frac{2}{3}P^C_{EW}  \to t_c~,~~~e_c - \frac{2}{3}PA_{EW} \to e_c~,
p_u - \frac{2}{3}P^C_{EW} \to p_u~,~~~pa_u -\frac{2}{3}PA_{EW} \to pa_u~.
\eeq

\begin{table}
\begin{center}
{\begin{tabular}{cccc} 
\hline\hline
Decay mode: &$B^0\to D^+_sD^-$ & $B^+\to D^+_s\bar D^0$ & $B_s\to D^+_sD^-_s$\\
\hline
Branching ratio: & $79\pm 7$~\cite{PDGup,Aubert:2006nm,Zupanc:2007pu} & $ 109 \pm 18$~\cite{PDGup,Aubert:2006nm} 
& $94^{+44}_{-42}$~\cite{CDFBsDsDs}\\
\hline\hline
\end{tabular}}
\caption{Charged averaged branching ratios in units of $10^{-4}$ 
for $\Delta S=1$ $B$ and $B_s$ decays into charm-anticharm.}
\label{tab:2}
\end{center}
\end{table}
We conclude this section by quoting in Table~\ref{tab:2} branching ratios for three of the 
five processes occurring in Eqs.~(\ref{5amps'}), including a very recent measurement 
of $\b(B_s\to D^+_sD^-_s)$ by the CDF Collaboration~\cite{CDFBsDsDs}.

\section{Expected hierarchy among graphical amplitudes}\label{hierarchy}

\subsection{$\Delta S=1$ decays}
\subsubsection{Ratio of two CKM factors}
Consider first the five $\Delta S=1$ amplitudes (\ref{5amps'}) for  $B$ and $B_s$ decays 
into pairs of charmed pseudoscalar mesons. Each amplitude involves a dominant term 
with a CKM factor $V^*_{cb}V_{cs}$ and a much smaller term involving $V^*_{ub}V_{us}$.
The CKM suppression of the second amplitude, which is often being neglected, 
is~\cite{CKMfitter}:
\beq\label{tinyCKM}
\frac{|V^*_{ub}V_{us}|}{|V^*_{cb}V_{cs}|}=(0.40\pm 0.05)\lambda^2/(1-\lambda^2)=0.021\pm 0.003~.
\eeq

\subsubsection{The amplitudes $t_c$ and $e_c$}
The large CKM factor $V^*_{cb}V_{cs}$ multiplies two  amplitudes, a dominant term $t_c$ 
and a smaller exchange contribution $e_c$ accompanied by an EWP-annihilation contribution $PA_{EW}$ [see (\ref{EWP-trans})].
The latter two terms are suppressed by $\Lambda_{\rm QCD}/m_B$ as they involve  
the interaction of a light  spectator quark. Thus, we expect the 
two decay modes $B_s\to D^+D^-$ and $B_s\to D^0\bar D^0$ governed by $e_c$ to have
branching ratios much smaller~\cite{Blok:1997yj} than those quoted in Table \ref{tab:2} for $B^0\to D^+_sD^-, B^+\to D^+_s\bar D^0$ and $B_s\to D^+_sD^-_s$ which are dominated by $t_c$.
The corresponding small ratios of branching ratios, e.g. $\b(B_s\to D^+D^-)/\b(B^0\to D^+_sD^-)$, 
would determine $|e_c/t_c|^2$.

A reasonable although not precise estimate for $|e_c/t_c|$ may proceed as follows. 
Consider the two processes $B^0\to D^-\pi^+$ and $B^0\to D^-_sK^+$, both of which originate 
in the quark subprocess $\bar b\to \bar c u\bar d$. While the first decay is governed by a 
color-favored tree amplitude with a small exchange contribution, the second one obtains only 
a contribution from an exchange amplitude. Drawing a 
parallel between these amplitudes and the corresponding amplitudes in $\bar b\to\bar c c\bar s$, 
and using~\cite{PDGup} $\b(B^0\to D^-\pi^+)=(26.8 \pm 1.3)\times 10^{-4}, \b(B^0\to D^-_sK^+)=
(2.8 \pm 0.5)\times 10^{-5}$, we estimate
\beq
\frac{|e_c|}{|t_c + e_c|} \sim \sqrt{\frac{\b(B^0\to D^-_sK^+)}{\b(B^0\to D^-\pi^+)}}= 0.102 \pm 0.009~.
\eeq
In this crude approximation this would imply 
\beq\label{ec/tc}
0.093\pm 0.008 \le \frac{|e_c|}{|t_c|} \le 0.114 \pm 0.010~.
\eeq
We note three corrections which may affect this estimate:
\begin{enumerate}
\item The amplitude $e_c$ includes by definition a term $PA_{EW}$ (both terms require
an interaction of the spectator quark), while no such  term contributes to $B^0\to D^-_sK^+$. 
\item The exchange amplitude in $B^0\to D^-_sK^+$ involves $s\bar s$ popping in a $\bar cu$ 
system, whereas $e_c$ is described by $u\bar u$ or $d\bar d$ popping in a $\bar cc$ system.
\item $B^0\to D^-\pi^+$ is dominated by a purely a color-favored tree amplitude, while
$t_c$ involves also a smaller penguin term $p_{ct}$. See discussion below. 
\end{enumerate} 
These three differences are expected to affect (\ref{ec/tc}) by a factor which is hard to estimate.

In our discussion below we will make a conservative assumption
based on measurements for $\Delta S=0$ decays, 
\beq\label{ec/tc-bound}
\frac{|e_c|}{|t_c|} \approx \sqrt{\frac{\b(B^0\to D^0\bar D^0)}{\b(B^+\to D^+\bar D^0)}
\frac{\tau_+}{\tau_0}} \le 0.3~.
\eeq
This upper bound is obtained from the branching ratios quoted in 
Table~\ref{tab:1} for $B^0\to D^0\bar D^0$ and $B^+\to D^+\bar D^0$, for   
which $1\sigma$ upper and lower limits are used.
The two processes $B^0\to D^0\bar D^0$ and $B^+\to D^+\bar D^0$ are dominated by 
$e_c$ and $t_c$, respectively, while smaller contributions involving $V^*_{ub}V_{ud}$ 
have been  neglected. (See discussion below.)  It would be useful to compare the bound 
(\ref{ec/tc-bound}) and the crude estimate (\ref{ec/tc}) with direct 
measurements of $|e_c/t_c|$ in ratios of branching ratios for $\Delta S=1$ decays including 
$\b(B_s\to D^+D^-)/\b(B^+\to D^+_s\bar D^0)$ and $\b(B_s\to D^0\bar D^0)/\b(B^0\to D^+_sD^-)$. 
Using the averaged measured $\b(B^+\to D^+\bar D^0)$ in Table I to normalize $|t_c|^2$, 
Eq.~(\ref{ec/tc}) would imply $\b(B^0\to D^+_sD^-_s) = 
(4.0^{+1.8}_{-1.4})\times 10^{-6}$, about an order of magnitude smaller than  
the current upper limit on this branching ratio (see Table I.)

\subsubsection{The ratio $|V^*_{cb}V_{cd}p_{ct}/T|$ in $t_c$}
The amplitude $t_c$ consists of a 
combination of a genuine color-favored tree amplitude and a smaller loop-suppressed penguin amplitude, $p_{ct}$, with $t$ and $c$ quarks in the loop 
[see Eq.~(\ref{t_c})]. It is difficult to obtain a good estimate for the ratio of these two amplitudes,
the sum of which contributes to both $\Delta S=0$ and $\Delta S=1$ decays. A QCD loop factor 
$[\alpha_s(m_b)/12\pi]\ln(m^2_t/m^2_c)$ characterizing the suppression of $V^*_{cb}V_{cd}p_{ct}$ 
relative to $T$ (or a typical Wilson coefficient for penguin operators~\cite{Buchalla:1995vs}) is 
of order five percent.  A dynamical enhancement by a factor of four to six relative to a  
QCD loop factor has been measured for the penguin-to-tree ratio in 
$B^0\to \pi^+\pi^-$ ~\cite{Chiang:2004nm,Gronau:2004ej}. We will permit a similar enhancement 
in $B^0\to D^+D^-$, thus allowing $|V^*_{cb}V_{cd}p_{ct}|$ to be at most as large as $0.3|T|$,
\beq\label{pct/T}
\frac{|V^*_{cb}V_{cd}p_{ct}|}{|T|} \le 0.3~.
\eeq

\subsubsection{Factorization of $(V_{cs}/V_{cd})T$ in $B^0\to D^+_s D^-$ and 
$B^+\to D^+_s\bar D^0$}
The tree amplitude is expected to factorize within a 
reasonable approximation into a product of a $B\to D$ form factor and the $D_s$ meson 
decay constant. While this approximation cannot be justified by the heavy $b$ quark limit of 
QCD (which can only be applied when a $B$ meson decays into two energetic mesons~\cite{Bjorken:1988kk,Beneke:2000ry}), it holds to leading order in $1/N_c$ in the large 
$N_c$ limit~\cite{Bauer:1986bm}. Early factorization tests of this kind, implicitly
neglecting  a $p_{ct}$ contribution, have been proposed
and studied in Ref.~\cite{Bortoletto:1990np,Rosner:1990xx}.

We will now update a factorization test for $(V_{cs}/V_{cd})T$ in $B^0\to D^+_sD^-$ 
and $B^+\to D^+_s\bar D^0$
by relating the branching ratios for these two processes given in Table~\ref{tab:2} 
to the above quoted branching ratio for $B^0\to D^-\pi^+$. The latter  
has been  accounted rather well by factorization~\cite{Beneke:2000ry}, up to a small 
contribution from an exchange amplitude.
The dominant contributions of the isosinglet amplitudes $(V_{cs}/V_{cd})T$ and 
$V^*_{cb}V_{cs}p_{ct}$
to the decay rates of $B^+\to D^+_s\bar D^0$ and $B^0\to D^+_sD^-$
are expected to be equal in the isospin symmetry limit. Thus we will 
use the weighted average of these two branching ratios, correcting $\b(B^+\to D^+_s\bar D^0)$
by the lifetime ratio $\tau_0/\tau_+$:
\beq\label{DsD}
\tilde\b(B\to D^+_s\bar D)  =   (82.4 \pm 6.5)\times 10^{-4}~.
\eeq
This implies the following  ratio of measured branching ratios,
\beq\label{ratio-factor}
\frac{\tilde\b(B\to D^+_s\bar D)}{\b(B^0\to D^-\pi^+)} = 3.07 \pm 0.28~.
\eeq

Assuming factorization for the processes in the numerator and denominator and taking
$V_{cs}/V_{ud}=1$, one would expect this ratio to be given by~\cite{Rosner:1990xx}
\beq
\frac{\tilde\b(B\to D^+_s\bar D)}{\b(B^0\to D^-\pi^+)}  = \frac{f^2_{D_s}}{f^2_{\pi}}
\frac{F^2_V(\omega_{D_s})}{F^2_V(\omega_{\pi})}
\frac{[(1+\sqrt{\zeta_D})^2 - \zeta_{D_s}]^2}{[(1+\sqrt{\zeta_D})^2 - \zeta_{\pi}]^2}
\frac{p_{D_s}}{p_{\pi}}~,
\eeq
where
\beq
\omega_x\equiv \frac{m^2_B +m^2_D -m^2_x}{2m_Bm_D}~,~~~
\zeta_x\equiv \frac{m^2_x}{m^2_B}~.
\eeq
Here $f_{D_s},~f_{\pi}$ and $p_{D_s},~p_{\pi}$ are corresponding decay constants and
momenta in the $B$ meson rest frame, while $F_V$ is the $B\to D$ vector form factor.
Taking a linear form factor~\cite{Abe:2001yf}, $F_V(\omega)=
F_V(1)[1-(0.69\pm 0.14)(\omega-1)]$, and  using~\cite{Rosner:2008yu} $f_{\pi}=130.4 \pm 0.2$ 
MeV,~$f_{D_s}=273\pm 10$ MeV,  with meson masses and momenta in the $B$ rest frame listed 
in~Ref.~\cite{PDGup}, one finds
\beq\label{factor}
\frac{\tilde\b(B\to D^+_s\bar D)}{\b(B^0\to D^-\pi^+)}  = 4.42^{+0.74}_{-0.57}~.
\eeq

Comparing the factorization calculation with the experimental ratio (\ref{ratio-factor}) 
we note that the factorization result is on the high side, showing a discrepancy 
of $2.1\sigma$ relative to experiment.  We do not expect a very good agreement here 
because of two corrections, each of which may be about $30\%$ in amplitude:
\begin{enumerate}
\item A penguin term $V^*_{cb}V_{cs}p_{ct}$ contributing to the numerator but not to the 
denominator. As mentioned, this term could be as large as $0.3T$ and could interfere 
destructively with $T$, leading to a ratio smaller than (\ref{factor}) by up to a factor of two.
\item Nonfactorizable $1/N_c$ corrections to $T$ contributing to both numerator and 
denominator. These terms are also expected to lead to corrections around $30\%$ in the 
amplitude.
\end{enumerate}
 
 \subsection{$\Delta S=0$ decays}
 \subsubsection{Ratio of two CKM factors}
We now turn to the $\Delta S=0$ decay amplitudes given in Eqs.~(\ref{5amps}). These amplitudes
involve the two CKM factors $V^*_{cb}V_{cd}$ and $V^*_{ub}V_{ud}$ which are of 
comparable order $\lambda^3$, with a ratio~\cite{CKMfitter}
\beq\label{Ru}
\frac{|V^*_{ub}V_{ud}|}{|V^*_{cb}V_{cd}|} = 0.40\pm 0.05~.
\eeq
In the U-spin symmetry limit these CKM factors multiply the same hadronic
amplitudes occurring in $\Delta S=1$ decays.
The somewhat larger CKM factor $V^*_{cb}V_{cd}$ multiplies a 
dominant term $t_c$ and an exchange contribution $e_c$ which is expected to be
much smaller. This leads to a large suppression of $\b(B^0\to D^0\bar D^0)$ and 
$\b(B^0\to D^+_sD^-_s)$ relative to $\b(B^0\to D^+D^-), \b(B^+\to D^+\bar D^0)$ and 
$\b(B_s\to D^+D^-_s)$~\cite{Blok:1997yj,Datta:2003va}. 

\subsubsection{Factorization of $T$ in $B^0\to D^+D^-$ and $B^+\to D^+\bar D^0$}
The hadronic amplitude $V^*_{cb}V_{cd}t_c$ consists of a dominant 
term $T$, which is factorizable in the large $N_c$ 
limit, and a sub-dominant term $V^*_{cb}V_{cd}p_{ct}$, which may be as large as about  $0.3T$.
Using notations as above, factorization of $T$ implies that the contribution of this amplitude to 
the rates for $B^0\to D^+D^-$ and $B^+\to D^+\bar D^0$ are equal and are given by 
\bea
&& \frac{\b(B^0\to D^+D^-)_T}{\b(B^0\to D^-\pi^+)}  = 
\frac{\b(B^+\to D^+\bar D^0)_T}{\b(B^0\to D^-\pi^+)} 
\nonumber\\
& & = \frac{\lambda^2}{1-\lambda^2}\frac{f^2_D}{f^2_{\pi}}
\frac{F^2_V(\omega_{D})}{F^2_V(\omega_{\pi})}
\frac{[(1+\sqrt{\zeta_D})^2 - \zeta_D]^2}{[(1+\sqrt{\zeta_D})^2 - \zeta_{\pi}]^2}
\frac{p_{D}}{p_{\pi}} = 0.136^{+0.022}_{-0.018}~.
\eea
We have used a value~\cite{Rosner:2008yu}  $f_D=205.8\pm 8.9$ MeV.
Using the measured branching ratio, $\b(B^0\to D^-\pi^+)=(26.8\pm 1.3)\times 10^{-4}$, 
this implies
\beq
\b(B^0\to D^+D^-)_T= \b(B^+\to D^+\bar D^0)_T = (3.64^{+0.62}_{-0.52})\times 10^{-4}~.
\eeq
This result is in agreement, well within $1\sigma$, with  the branching ratios measured 
by BaBar and Belle for $B^+\to D^+\bar D^0$. It  deviates from the BaBar and Belle 
measurements of $\b(B^0\to D^+D^-)$ by $1.0\sigma$ and $2.8\sigma$, respectively. 
(See Table \ref{tab:1}.) As we mentioned, deviations at this level are expected due to 
a term $V^*_{cb}V_{cd}p_{ct}$ and $1/N_c$ corrections. 

\subsubsection{The amplitude $p_u$ and the smaller terms, $pa_u, e_u, a_u$}
The CKM factor $V^*_{ub}V_{ud}$ in $\Delta S=0$ decay amplitudes multiplies four 
hadronic terms, $p_u, pa_u, e_u$ and $a_u$. The QCD penguin amplitude 
$p_u\equiv p_{ut}$, involving $t$ and $u$ quarks in the loop, is expected to have a 
magnitude comparable to that of $p_{ct}$ multiplying $V^*_{cb}V_{cd}$. Since we anticipate 
$|V^*_{cb}V_{cd}p_{ct}| \le 0.3|T|$ Eq.~(\ref{Ru}) implies 
\beq\label{pu/tc}
\frac{|V^*_{ub}V_{ud}p_u|}{|V^*_{cb}V_{cd}t_c|} \simeq  
\frac{|V^*_{cb}V_{cd}p_{ct}|}{|T+V^*_{cb}V_{cd}p_{ct}|} 
\frac{|V^*_{ub}V_{ud}|}{|V^*_{cb}V_{cd}|}
\le \frac{0.3}{1-0.3}\frac{|V^*_{ub}V_{ud}|}{|V^*_{cb}V_{cd}|} 
= 0.2~.
\eeq
This upper bound assumes a worst-case scenario of destructive interference between $T$ and 
$V^*_{cb}V_{cd}p_{ct}$ in $t_c$, as indicated by the factorization prediction (\ref{factor}) which
is larger than the corresponding experimental ratio (\ref{ratio-factor}).

The other three amplitudes, $pa_u, e_u$ and $a_u$, involving an interaction of a spectator
quark, are expected to be suppressed by $\Lambda_{\rm QCD}/m_B$.  For instance, 
the weak annihilation amplitude $a_u$ factorizes at leading order in  $1/N_c$,
\begin{eqnarray}\label{Aufact}
a_u = \frac{G_F}{\sqrt2} (C_1 + \frac{C_2}{N_c}) \langle 0|\bar b
\gamma_\mu
(1-\gamma_5) u|B^+\rangle \langle D^+\bar D^0 |\bar u \gamma_\mu d|0
\rangle + O(1/N_c^2)~.
\end{eqnarray}
The $|0\rangle\to |D^+\bar D^0\rangle $ matrix element of the $I=1$ vector current is  
parameterized by one form factor,
\begin{eqnarray}
\langle D^+(p_D)\bar D^0(p_{\bar D}) |\bar u \gamma_\mu d|0\rangle =  
f_+^{(I=1)}(q^2)
(p_D - p_{\bar D})_\mu~.
\end{eqnarray}
Combining the two factors in Eq.~(\ref{Aufact}), one finds that the leading term in $a_u$  
is proportional to the isospin breaking $D$ meson mass difference which is 
negligibly  small,
\begin{eqnarray}
a_u = \frac{G_F}{\sqrt2} (C_1 + \frac{C_2}{N_c}) f_B f_+^{(I=1)}
(m_B^2) (m_{D^+}^2 - m_{D^0}^2) + O(1/N_c^2)~.
\end{eqnarray}
This implies that $a_u$ is dominated by nonfactorizable  
contributions including initial state gluon emission.

Since $pa_u$ is also suppressed by a QCD loop factor, and $e_u$ 
and $a_u$ are suppressed by requiring a popping of a heavy $c\bar c$ pair, we will assume
\beq\label{neglect}
 |pa_u|, |e_u|, |a_u| \ll |p_u|~.
\eeq
Thus, in the subsequent analysis we will neglect these very small amplitudes.
[For the same argument, the interaction of a spectator quark and $c\bar c$ popping,
we have neglected in Section \ref{DS=1} the two EWP contributions, $PE_{EW}(c)$ and 
$PA_{EW}(c)$.] 

\section{Reiterating $B\to D\bar D$ decays and a bound on $r$}\label{DS=0hier}
Neglecting the very small amplitudes in (\ref{neglect}), Eqs.~(\ref{5amps}) (\ref{r,del}) 
and the last of Eqs.~(\ref{3half})  become:
\bea\label{5amps-approx}
a.~~~~~A(B^0 \to D^+D^-) &=& V^*_{cb}V_{cd}\,(t_c + e_c)  + V^*_{ub}V_{ud}\,p_u ~,
\nonumber\\
b.~~~~~~A(B^0 \to D^0 \bar D^0) &=& -V^*_{cb}V_{cd}\,e_c~,
\nonumber\\
c.~~~~~A(B^+ \to D^+\bar D^0) &=& V^*_{cb}V_{cd}\,t_c  + V^*_{ub}V_{ud}\,p_u~,
\nonumber\\
d.~~~~~A(B^0 \to D^+_s D^-_s) &=& V^*_{cb}V_{cd}\,e_c~,
\nonumber\\
e.~~~~~A(B_s \to D^+D^-_s) &=& V^*_{cb}V_{cd}\,t_c  + V^*_{ub}V_{ud}\,p_u~,
\eea
\beq\label{r-delta-A32}
r  =  \frac{|V^*_{ub}V_{ud}|}{|V^*_{cb}V_{cd}|}\frac{|p_u|}{|t_c + e_c|}~,~~~~~~
\delta = {\rm arg}\left(-\frac{p_u}{t_c + e_c}\right)~,~~~~~~
A_{1,\3half}  =  0~.
\eeq
Using the two upper limits (\ref{ec/tc-bound}) and (\ref{pu/tc}) we find
\beq\label{r-bound}
r \le 0.3~.
\eeq
We consider this a conservative upper bound, as it allows for a worst-case scenario
of two destructive interference terms in the denominator of $r$ and for a large 
enhancement of the penguin amplitude in its numerator.

We will now discuss the rate and asymmetry measurements in $B\to D\bar D$ in light of 
the expressions (\ref{5amps-approx}), which are expected to hold to a very good 
approximation. 
First, we note that CP asymmetries in $B^0\to D^0\bar D^0$ and $B^0\to
D^+_sD^-_s$  vanish in this approximation because the amplitudes for these
processes involve a single CKM factor  $V^*_{cb}V_{cd}$. [This justifies the
discussion below Eq.~(\ref{num-triangle} ).] The decay rates for these two
processes, which are dominated by an amplitude $e_c$, are equal in the SU(3)
limit. A small decay rate difference between $B^0\to D^0\bar D^0$ and $B^0\to
D^+_sD^-_s$ is expected due to two effects working in opposite directions,
$u\bar u$ versus $s\bar s$ popping on the one hand and exclusive production of
$D^0\bar D^0$ versus $D^+_sD^-_s$ on the other.

Second, as mentioned, neglecting $e_u$ and $a_u$ leads to $\Delta I=1/2$ in
$B\to D\bar D$ implying the triangle amplitude relation (\ref{triangle}).  This relation 
was shown to be violated at $1.8\sigma$ by branching
ratios and asymmetries measured by Belle.
The Standard Model $\Delta I=3/2$ amplitude $e_u +a_u$ is too small to account
for an observable violation.

The Belle asymmetries by themselves also show two unexpected features (see
Table \ref{tab:1}), which are not shared by the BaBar measurements:
\begin{enumerate}
\item The above upper bound on $r$ and the second Eq.~(\ref{SACP-exact})
imply a theoretical upper limit on the direct asymmetry in $B^0\to D^+D^-$,
\beq
|A_{CP}(B^0\to D^+D^-)| \le \frac{2r}{1+r^2} \le 0.55~.
\eeq 

The value measured by Belle is larger than this upper limit by $1.5\sigma$.

\item The asymmetries in $B^0\to D^+D^-$ and $B^+\to D^+\bar D^0$ are expected
to be equal, up to second order corrections from an interference of the small
amplitudes $e_c$ and $p_u$. In contrast to this expectation, the Belle
asymmetry in $B^0\to D^+D^-$ is positive and large while the one in $B^+\to
D^+\bar D^0$ is very small. The difference between the two asymmetries involves
a statistical significance of $3.6\sigma$. The Standard Model interference of $e_c$ 
and $p_u$ and the amplitude $pa_u$, which we neglected in (\ref{5amps-approx}a), 
are too small to account for this
large difference between the two CP asymmetries. 
\end{enumerate}

\section{Determining $\gamma$}
The amplitude for $B^0\to D^+D^-$ given in (\ref{5amps}a) or (\ref{5amps-approx}a)
and a suitably chosen $\Delta S=1$ amplitude in (\ref{5amps'}) provide a sufficient number 
of observables for determining the weak phase $\gamma$ in the flavor SU(3) limit. 
This method has been proposed in Refs.~\cite{Fleischer:1999nz} and~\cite{Datta:2003va}.
Here we wish to recapitulate this method, showing that the determination 
of $\gamma$ in this way is very sensitive to uncertainties in  SU(3) breaking.

The two asymmetries $S(D^+D^-)$ and $A_{CP}(D^+D^-)$ are given in Eqs.~(\ref{SACP-exact})
in terms of the three parameters $r, \delta$ and $\gamma$. We are assuming $\beta = (21.5\pm 1.0)^\circ$. A third equation for these parameters is provided by the ratio of CP-averaged
decay rates for $B^0\to D^+D^-$ and its U spin counterpart $B_s \to D_s^+D_s^-$. Alternatively, 
one may use instead of the latter process the decay mode $B^0\to D^+_sD^-$ or 
$B^+\to D^+_s\bar D^0$. In this case  one would have to  estimate the  effect of the exchange 
amplitude $e_c$ contributing to $B^0\to D^+D^-$ but not to the latter two $\Delta S=1$  
decay processes. [See discussion above including the upper bound (\ref{ec/tc-bound}).]

Focusing our attention on the U spin pair $B^0\to D^+D^-$ and $B_s \to D^+_sD^-_s$,  
we define an experimentally measured ratio of CP-averaged decay rates,
\beq
R \equiv \left (\frac{V_{cs}}{V_{cd}}\right)^2
\frac{\bar\Gamma(B^0\to D^+D^-)}{\bar\Gamma(B_s\to D_s^+D_s^-)} =
\frac{1-\lambda^2}{\lambda^2}\frac{\b(B^0\to D^+D^-)}{\b(B_s\to D^+_sD^-_s)}
\frac{\tau_s}{\tau_0}~,
\eeq
where~\cite{Barberio:2007cr} $\tau_s/\tau_0= 0.939\pm 0.021$ is the ratio of $B_s$ 
and $B^0$ lifetimes. Neglecting the second term in (\ref{5amps'}a) suppressed by the tiny 
CKM factor (\ref{tinyCKM}), and introducing an SU(3) breaking parameter $\xi$ for the ratio 
of $t_c+e_c$ amplitudes in $B_s \to D^+_sD^-_s$ and $B^0\to D^+D^-$, one obtains:
\beq\label{xi2R} 
\xi^2R = 1 + 2r\cos\delta\cos\gamma + r^2~.
\eeq 
Thus, for a given value of $\xi$, the three observables $S(D^+D^-), A_{CP}(D^+D^-)$ and $R$ in 
Eqs.~(\ref{SACP-exact}) and (\ref{xi2R}) enable a determination 
of $r, \delta$ and $\gamma$ up to discrete ambiguities.

In order to obtain an analytic solution for $\gamma$, and to overcome two of its 
four discrete ambiguities,  it is convenient to introduce another observable in 
$B^0\to D^+D^-$~\cite{Gronau:2002qj} (see also~\cite{Datta:2003va}),
\beq
D \equiv \frac {2{\rm Re}(\lambda_{D^+D^-})}{1 + |\lambda_{D^+D^-}|^2}~,
\eeq
obeying with the two asymmetries,
\beq\label{ASD}
A^2_{CP} + S^2 + D^2 = 1~.
\eeq
Expressing the new observable in terms of $r, \delta$ and $\gamma$, 
\beq\label{D}
D = \frac{\cos 2\beta + 2r\cos\delta\cos(2\beta+\gamma) + r^2\cos 2(\beta +\gamma)}
{1 + 2r\cos\delta\cos\gamma +r^2}~,
\eeq
we note that this quantity is positive for $\beta = (21.5\pm 1.0)^\circ, r\le 0.3$ 
[as required by (\ref{r-bound})], and for arbitrary values of $\delta$ and $\gamma$.
This information on the sign of $D$, which remains undetermined by the two asymmetries 
using (\ref{ASD}), avoids two discrete ambiguities in $\gamma$.  

Eqs.~(\ref{SACP-exact}) and (\ref{xi2R}) and the positivity of $D$ imply
the following equation for $\gamma$ in terms of $A_{CP}, S$ and $\xi^2R$~\cite{Datta:2003va}:
\beq
\frac{+\sqrt{1 - A^2_{CP} - S^2}\cos 2(\beta +\gamma) - S\sin 2(\beta +\gamma) - 1}
{\cos 2\gamma -1} = \frac{1}{\xi^2R}~.
\eeq
The plus sign in front of the square root follows from the positivity of $D$.

\begin{figure}[t!]
\centering{\includegraphics[width=9.5cm]{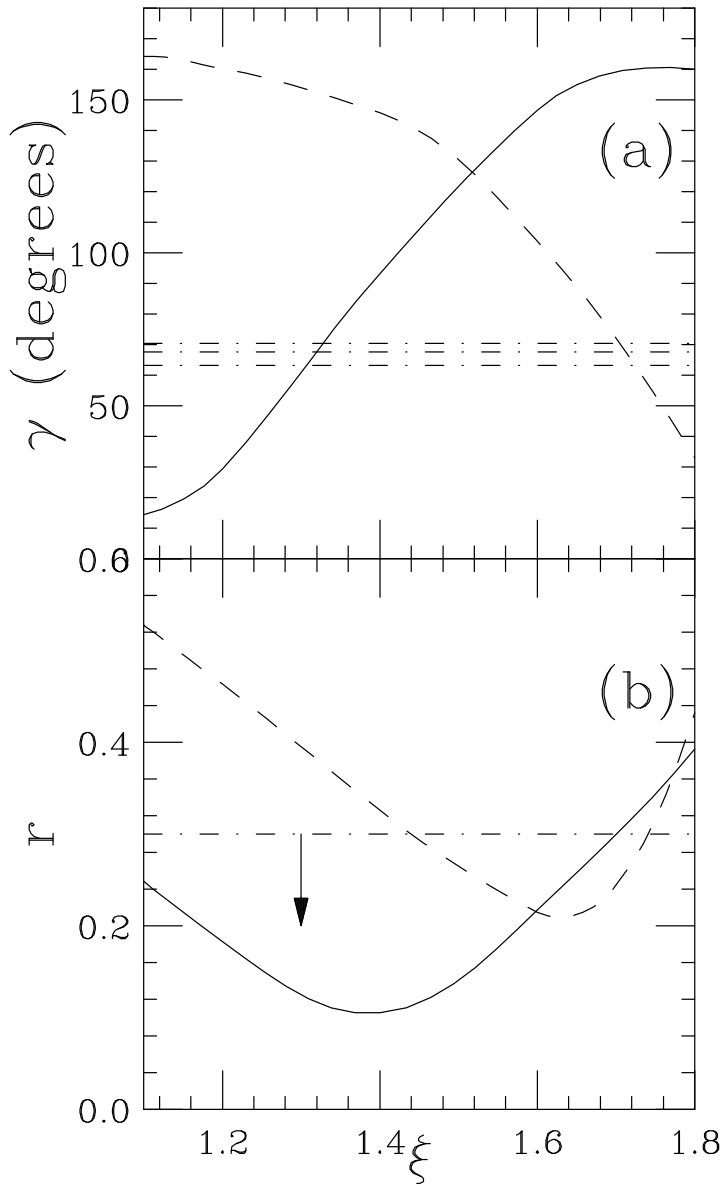}}
\caption{Behavior of solutions as a function of parameter $\xi$ describing SU(3)
violation.  Solid curves denote values obtained using BaBar central values
as input; dashed curves based on BaBar--Belle average central values.    
(a) Weak phase $\gamma$.  Horizontal
dashed lines denote HFAG central and $\pm 1 \sigma$ values of $\gamma$.
(b) Parameter $r= A_u/A_c$ describing ratio of amplitudes.  Horizontal
dot-dashed line denotes upper bound $r \le 0.3$ described in text.}
\label{fig:rgamma}
\end{figure}

In order to demonstrate  the high sensitivity of determining $\gamma$ to the value of the SU(3)
breaking parameter $\xi$, we plot in Fig.~2(a) the dependence of $\gamma$ on $\xi$ for
two sets of values for  $A_{CP}, S$ and $R$: \\
(1) BaBar central values, $A_{CP}=-0.11, S=-0.54, R=0.52$ (solid curve).\\ 
(2) Central values of the averages of BaBar and Belle, $A_{CP}=0.37, S=-0.75, R=0.39$
(dashed curve).\\
We do not use the Belle central values because they are non-physical, violating the 
inequality $S^2 + A_{CP}^2 \le 1$. Also shown in the plot is a band describing the currently 
allowed $1\sigma$ range for $\gamma$~\cite{CKMfitter}, $\gamma= (67.6^{+2.8}_{-4.5})^\circ$.
Fig.~2(b) shows values of $r$ corresponding to solutions for $\gamma$ for case (1) (solid curve) 
and (2) (dashed curve). A solution obtained with too large values of $r$, $r>1.5$ in 
case (1) and $r> 0.6$ in case (2), is not shown.

The two curves of $\gamma$ for case  (1) and (2) cross the allowed band for $\gamma$ at 
$\xi=1.3$ and $\xi = 1.7$,
respectively. This difference in the values of the SU(3) breaking parameter follows 
from the large experimental errors in the $B^0\to D^+D^-$ measurements.
The slopes of the two curves at the above points are quite steep, implying in both cases a high 
theoretical sensitivity of the determined value of $\gamma$ to the assumed value of $\xi$. 
For instance, the dashed curve increases approximately linearly from $\gamma=65^\circ$ at 
$\xi= 1.7$ to $\gamma = 105^\circ$ at $\xi =1.6$.  Thus, assuming perfect measurements 
for the branching ratio and asymmetries in $B^0\to D^+D^-$, an uncertainty of $10^\circ$ in 
$\gamma$ would require knowing $\xi$ to better than $2\%$.
A somewhat lower sensitivity has been noted in the second paper in 
Ref.~\cite{Fleischer:1999nz}.

The parameter $\xi$, representing SU(3) breaking in the ratio of $t_c+e_c$ amplitudes in
$B_s\to D^+_sD^-_s$ and $B^0\to D^+D^-$ including a small calculable phase space effect, 
involves theoretical uncertainties from several
sources:
\begin{itemize}
\item As mentioned, $t_c$ include a dominant tree amplitude, for which the leading
term in a $1/N_c$ expansion factorizes. SU(3) breaking in the factorizable term is  given by 
a product of a measured ratio of decay constants\cite{Rosner:2008yu}, 
$f_{D_s}/f_D=1.33 \pm 0.07$, and a ratio of  form factors 
$F_{B_s\to D_s}(m^2_{D_s})/F_{B\to D}(m^2_D)$. The latter ratio is estimated at $\sim 1.05$
in  a chiral SU(3) perturbation expansion, at leading order in the heavy $b$ and $c$ quark  
masses $m_H$~\cite{Jenkins:1992qv}. 
However, a complete analysis of all ${\cal O}(1/m_H)$ terms shows that simultaneous violation of 
both chiral and heavy quark symmetries can be as large as $30\%$~\cite{Boyd:1995pq}.
The form factor ratio can eventually be determined from $B^0\to D^-\ell^+\nu$ measured in 
$e^+e^-$ collisions on the $\Upsilon(4S)$~\cite{Abe:2001yf} and
$B_s \to D^-_s\ell^+\nu$ accessible to the LHCb collaboration working at the LHC. 
Thus, SU(3) breaking in the tree amplitude 
involves  nonfactorizable $1/N_c$ corrections and  uncertainties in ratios of form factors and 
decay constants combining to a total of at least $30\%$. 
\item A penguin contribution $V^*_{cb}V_{cd}p_{ct}$ in $t_c$ is not expected to 
factorize, which introduces an uncontrollable SU(3) breaking correction in this contribution
at a level of $30\%$. 
\item We have already discussed uncertainties in the magnitude of  
$e_c$. $30\%$ uncertainties due to SU(3) breaking corrections in this amplitude and in 
$p_{ct}$ mentioned above translate through (\ref{ec/tc-bound}) and (\ref{pct/T}) into two 
uncertainties in SU(3) 
breaking in $t_c +e_c$, each of which is at a level of $10\%$.  
\end{itemize}
In view of these combined uncertainties, a precision of $10\%$ in $\xi$ is unachievable.
This implies a very large intrinsic theoretical uncertainty of at least $50^\circ$ in 
determining $\gamma$ as demonstrated in Fig.~2(a).

The origin of the high sensitivity to SU(3) breaking can be traced back to the
way flavor SU(3) symmetry is being applied here for a determination of $\gamma$. 
In this method an assumption of SU(3) symmetry is used to normalize the {\em dominant}
amplitude in $B^0\to D^+D^-$ in terms of the measured amplitude for 
$B_s\to D^+_sD^-_s$. In contrast, the error in $\gamma$ introduced by an uncertainty 
in SU(3) breaking is expected to be small in cases where a {\em small} penguin 
amplitude in $\Delta S=0$ decays is normalized by a measurable SU(3) related 
$\Delta S=1$ decay amplitude. Two cases, where this has been demonstrated, are
$B^0\to \pi^+\pi^-$~\cite{Gronau:2007af}, where the penguin amplitude is 
subdominant, and most prominently $B^0\to\rho^+\rho^-$~\cite{Beneke:2006rb} 
in which the penguin amplitude is very small.

\section{Conclusion}
We have studied the effect of a small $\bar b\to \bar uu\bar d$ amplitude in
$B\to D\bar D$ decays. While this amplitude violates $\Delta I=1/2$, we have
argued that this violation is expected to be very small.  
Considering the Belle measurements, we pointed out a violation at a level of 
$1.8\sigma$ of a $\Delta I=1/2$ amplitude relation, and an 
inconsistency at a level of $3.6\sigma$ 
between CP asymmetries in $B^0\to D^+D^-$ and $B^+\to D^+\bar D^0$.
No such inconsistency has been observed by the BaBar collaboration.
If these discrepancies persist they would have to be associated
with a New Physics contribution to $\bar b\to \bar uu\bar d$. 

Using conservative considerations within the CKM framework, we obtained a 
model-independent upper bound, $r\equiv A_u/A_c\le 0.3$, for the ratio of 
$\bar b\to \bar uu\bar d$ and $\bar b\to \bar cc\bar d$ amplitudes in $B^0\to
D^+D^-$.  This implies an upper limit, $|A_{CP}| \le 0.55$, for the direct CP
asymmetry in this process.  The Belle asymmetry is $1.5\sigma$ larger than
this bound, while the BaBar measurement is well below this upper limit.

U spin symmetry has been applied for obtaining relations between amplitudes and
CP asymmetries for $\Delta S=0$ decays of $B$ and $B_s$ to pairs of
charm-anticharm mesons and amplitudes and asymmetries in corresponding $\Delta
S=1$ decays.  We have shown that, while these relations are not useful for a
precise determination of $\gamma$ in $B^0\to D^+D^-$ in the presence of small
uncertainties in SU(3) breaking, they may provide important information about
small contributions to the latter process. 

For instance, the amplitude $e_c$ may account for a difference between the
decay rates of $B^0\to D^+D^-$ and $B^+\to D^+\bar D^0$, and for a small
difference between CP asymmetries in these decays.  This amplitude dominates
the $\Delta S=0$ decays, $B^0\to D^0\bar D^0,  B^0\to D^+_sD^-_s$, 
and the $\Delta S=1$ decays $B_s\to D^+D^-, B_s\to D^0\bar D^0$, and may be 
extracted from branching ratios of these processes.
Using Eq.~(\ref{ec/tc}) one would  predict $\b(B^0\to D^+_sD^-_s) = 
(4.0^{+1.8}_{-1.4})\times 10^{-6}$, whereas (\ref{ec/tc-bound}) only implies 
$\b(B^0\to D^+_sD^-_s) \le 3\times 10^{-5}$, close to the upper limit  in 
Table I~\cite{Zupanc:2007pu}. Similarly, using (\ref{DsD}) $\tilde\b(B\to
D^+_s\bar D) = (82.4\pm 6.5)\times 10^{-4}$ Eq.~(\ref{ec/tc}) would imply
$\b(B_s\to D\bar D)=(0.9^{+0.4}_{-0.3})\times 10^{-4}$, while the conservative
upper bound (\ref{ec/tc-bound}) leads to a more modest prediction $\b(B_s\to
D\bar D)\le 7\times 10^{-4}$.

Thus, a sensitivity of $1\times 10^{-4}$ in  $\b(B_s\to D^+D^-)$ and $\b(B_s\to
D^0\bar D^0)$, in comparison with $\tilde\b(B\to D^+_s\bar D)=(82.4\pm 6.5)
\times 10^{-4}$, can be used for obtaining precise information on the ratio
$|e_c/t_c|$ at a level of $0.1$. This precision is considerably more powerful
than the current upper bound (\ref{ec/tc-bound}) obtained from the $\lambda^2$
suppressed $\b(B^0\to D^0\bar D^0)$.  The above sensitivity may be achieved at
experiments carried by the LHCb collaboration working at the Large Hadron
Collider.  More precise measurement of the Cabibbo-favored decay branching ratio
$\b(B_s\to D^+_sD^-_s)$ 
than currently available may soon be achieved at the Tevatron. This would lead
to useful information on corrections to U spin symmetry relating this process
and $B^0\to D^+D^-$.

\bigskip
{\it Acknowledgments:} M.G. would like to thank the Enrico Fermi Institute at 
the University of Chicago for its kind and generous hospitality. We thank 
Pavel Krokovny, Shunzo Kumano, Manfred Paulini and Yoshi Sakai for 
useful communications and Sheldon Stone for comments about feasibility 
at LHCb.  This work was supported in part by 
the United States Department of Energy through Grant No.\ DE FG02 90ER40560.

\appendix
\section{SU(3) operator analysis}

In this Appendix we consider relations between the graphical amplitudes 
defined in Sections 4 and 5 and a flavor SU(3) analysis for the operators appearing in the 
low energy effective  Hamiltonian~(\ref{Heff}). These operators  have the following transformation
properties under flavor SU(3). Current-current operators: ${\cal O}_{1,2}^c \sim \overline{\mathbf 3}$,
${\cal O}_{1,2}^u \sim \overline{\mathbf 3}, {\mathbf 6}, \overline{\mathbf{15}}$,
QCD penguin operators: ${\cal O}_{3-6} \sim \overline{\mathbf 3}$, electroweak penguin operators: 
${\cal O}_{7-10} \sim \overline{\mathbf 3}, {\mathbf 6}, \overline{\mathbf{15}}$.
An explicit SU(3) decomposition of the Hamiltonian can be found e.g. in \cite{Gronau:1998fn}.

In  $\bar B\to D\bar D$ decays, the initial and final states transform like  ${\mathbf 3}$ and
${\mathbf 1}, {\mathbf 8}$, respectively, which implies that all these decay amplitudes 
can be expressed in terms of four SU(3) reduced matrix elements
$\langle\mathbf{1}|\!|\overline{\mathbf 3}|\!|{\mathbf 3}\rangle, \langle {\mathbf 8}|\!|
\overline{\mathbf 3}|\!|{\mathbf 3}\rangle, \langle{\mathbf 8}|\!|{\mathbf 6}|\!|{\mathbf 3}\rangle, 
\langle{\mathbf 8}|\!|\overline{\mathbf{15}}|\!|{\mathbf 3}\rangle$.
This agrees with the counting of independent amplitudes performed in Section 4 in 
terms of graphical amplitudes. 

The explicit expansion of all ten $\Delta S=0$ and $\Delta S=1$ amplitudes
in terms of reduced SU(3) matrix elements can be found, for example, in 
\cite{Grinstein:1996us}.
The $\Delta S=0$ amplitudes are 
\begin{eqnarray}
\left(
\begin{array}{c}
A(B^0\to D^+ D^-) \\
A(B^0\to D^0 \bar D^0) \\
A(B^+\to D^+ \bar D^0) \\
A(B^0\to D_s^+ D_s^-) \\
A(B_s\to D^+D_s^-) \\
\end{array}
\right) 
 = \left(
\begin{array}{cccc}
\frac{1}{\sqrt3} & -\frac{1}{\sqrt6} & 0 & -\frac12 \\
-\frac{1}{\sqrt3} & -\frac{1}{2\sqrt6} & -\frac12 & -\frac34 \\
0 & -\frac12\sqrt{\frac32} & -\frac12 & \frac34 \\
\frac{1}{\sqrt3} & \frac{1}{2\sqrt6} & -\frac12 & -\frac14 \\
0 & -\frac12 \sqrt{\frac32} & \frac12 & - \frac14 \\
\end{array}
\right)
\left(
\begin{array}{c}
\langle{\mathbf 1} |\! |\overline{\mathbf 3} |\!|{\mathbf 3}\rangle_d \\
\langle{\mathbf 8}|\! |\overline{\mathbf 3} |\!|{\mathbf 3}\rangle_d \\
\langle{\mathbf 8} |\! |{\mathbf 6} |\!|{\mathbf 3}\rangle_d \\
\langle{\mathbf 8} |\! |\overline{\mathbf{15}} |\!|{\mathbf 3}\rangle_d \\
\end{array}
\right)~.
\end{eqnarray} 

The corresponding $\Delta S=1$ amplitudes are given by the same transformation matrix,
\begin{eqnarray}
\left(
\begin{array}{c}
A(B_s\to D_s^+D_s^-) \\
A(B_s^0\to D^0 \bar D^0) \\
A(B^+\to D_s^+ \bar D^0) \\
A(B_s^0\to D^+ D^-) \\
A(B^0\to D_s^+ D^-) \\
\end{array}
\right) 
 = \left(
\begin{array}{cccc}
\frac{1}{\sqrt3} & -\frac{1}{\sqrt6} & 0 & - \frac12 \\
-\frac{1}{\sqrt3} & -\frac{1}{2\sqrt6} & -\frac12 & -\frac34 \\
0 & -\frac12\sqrt{\frac32} & -\frac12 & \frac34 \\
\frac{1}{\sqrt3} & \frac{1}{2\sqrt6} & -\frac12 & -\frac14 \\
0 & -\frac12\sqrt{\frac32} & \frac12 & -\frac14 \\
\end{array}
\right)
\left(
\begin{array}{c}
\langle{\mathbf 1} |\! |\overline{\mathbf 3} |\!|{\mathbf 3}\rangle_s \\
\langle{\mathbf 8} |\! |\overline{\mathbf 3} |\!|{\mathbf 3}\rangle_s \\
\langle{\mathbf 8} |\! |{\mathbf 6} |\!|{\mathbf 3}\rangle_s \\
\langle{\mathbf 8} |\! |\overline{\mathbf{15}} |\!|{\mathbf 3}\rangle_s \\
\end{array}
\right)~.
\end{eqnarray}
The subscript $q=d,s$ on the reduced matrix elements is a reminder that
they differ for $\Delta S=0,1$ weak Hamiltonians 
through their dependence on CKM matrix elements.

Comparing these expressions with the graphical expansions (\ref{5amps}) 
and (\ref{5amps'}), we find the following relations between 
SU(3) reduced matrix elements and graphical amplitudes (for $q=d,s$):
\begin{eqnarray}
\langle{\mathbf 1} |\! |\overline{\mathbf 3} |\!|{\mathbf 3}\rangle_q &=&
V_{cb}^* V_{cq} \frac{1}{\sqrt3} (t_c + 3e_c) +
V_{ub}^* V_{uq} \frac{1}{\sqrt3} (p_u+ 3pa_u +e_{u})~,  
\nonumber\\
\langle{\mathbf 8} |\! |\overline{\mathbf 3}|\!|{\mathbf 3}\rangle_q &=&
-V_{cb}^* V_{cq} \frac{1}{2\sqrt6}8 t_c 
-V_{ub}^* V_{uq} \frac{1}{2\sqrt6} (8p_u-e_u+3a_u)~,
\nonumber\\
\langle{\mathbf 8} |\! |{\mathbf 6} |\!|{\mathbf 3}\rangle_q &=&
V_{ub}^* V_{uq} \frac{1}{2} (e_u-a_u)~,
\nonumber\\
\langle{\mathbf 8} |\! |\overline{\mathbf{15}} |\!|{\mathbf 3}\rangle_q &=&
V_{ub}^* V_{uq} \frac{1}{2} (e_u+a_u)~.
\end{eqnarray}

As mentioned in Section 5, electroweak penguin (EWP) contributions 
introduce four new graphical amplitudes. This agrees with
the above counting of SU(3) reduced matrix elements. A complete expansion 
of EWP terms  in the ten processes (\ref{5amps}) and 
(\ref{5amps'}) in terms of graphical amplitudes defined in Section 5 is:
\begin{eqnarray}
P_{EW}(B^0\to D^+D^-) &=& 
  V^*_{tb}V_{td} (\frac23 P_{EW}^C - \frac13 PE_{EW(c)} + 
                  \frac23 PA_{EW} - \frac13 PA_{EW(c)})~,
\nonumber\\
P_{EW}(B^0\to D^0\bar D^0) &=& 
  V^*_{tb}V_{td} (-\frac23 PA_{EW} - \frac23 PA_{EW(c)})~,
  \nonumber\\
P_{EW}(B^+\to D^+\bar D^0) &=& 
  V^*_{tb}V_{td} (\frac23 P_{EW}^C + \frac23 PE_{EW(c)})~,
  \nonumber\\
P_{EW}(B^0\to D_s^+D_s^-) &=& 
  V^*_{tb}V_{td} (\frac23 PA_{EW} - \frac13 PA_{EW(c)})~,
  \nonumber\\
P_{EW}(B_s^0\to D^+D_s^-) &=& 
   V^*_{tb}V_{td} (\frac23 P_{EW}^C - \frac13 PE_{EW(c)})~.
\end{eqnarray} 
\begin{eqnarray}
P_{EW}(B_s\to D_s^+D_s^-) &=& 
  V^*_{tb}V_{ts} (\frac23 P_{EW}^C - \frac13 PE_{EW(c)} + 
                  \frac23 PA_{EW} - \frac13 PA_{EW(c)})~,
                  \nonumber\\
P_{EW}(B_s\to D^0\bar D^0) &=& 
  V^*_{tb}V_{ts} (-\frac23 PA_{EW} - \frac23 PA_{EW(c)})~,
  \nonumber\\
P_{EW}(B^+\to D_s^+\bar D^0) &=& 
  V^*_{tb}V_{ts} (\frac23 P_{EW}^C + \frac23 PE_{EW(c)})~.
  \nonumber\\
P_{EW}(B_s\to D^+D^-) &=& 
  V^*_{tb}V_{ts} (\frac23 PA_{EW} - \frac13 PA_{EW(c)})~,
  \nonumber\\
P_{EW}(B^0\to D_s^+D^-) &=& 
   V^*_{tb}V_{ts} (\frac23 P_{EW}^C - \frac13 PE_{EW(c)})~.
\end{eqnarray} 

Two of the EWP amplitudes can be related within a very good 
approximation to the amplitudes $e_u$ and $a_u$ appearing in Eqs.~(\ref{5amps})
and (\ref{5amps'}).  Neglecting the EWP operators ${\cal O}_{7,8}$
which have negligibly small Wilson coefficients, SU(3) symmetry implies,
\begin{eqnarray}\label{EWP2tree}
PE_{EW(c)} + PA_{EW(c)} &=& -\frac32 \frac{C_9+C_{10}}{C_1+C_2} (a_u + e_u)~,
\nonumber\\
PE_{EW(c)} - PA_{EW(c)} &=& \frac32 \frac{C_9-C_{10}}{C_1-C_2} (a_u - e_u)~.
\end{eqnarray}
The proof of these relations is based on operator relations between  
${\mathbf 6}$ and $\overline{\mathbf{15}}$ components of the EWP part of the
effective Hamiltonian and corresponding components of the tree 
part ($q=d,s$)~\cite{Gronau:1998fn,Gronau:2000az},
\begin{eqnarray}
{\cal H}^q_{EWP}({\overline{\mathbf 15}})= -\frac32 \frac{C_9+C_{10}}{C_1+C_2} 
\frac{V_{tb}^*V_{tq}}{V_{ub}^*V_{uq}} {\cal H}^q_T({\overline{\mathbf 15}})~,\quad
{\cal H}^q_{EWP}({\mathbf 6}) = \frac32 \frac{C_9-C_{10}}{C_1-C_2} 
\frac{V_{tb}^*V_{tq}}{V_{ub}^*V_{uq}} {\cal H}^q_T({\mathbf 6})~.
\end{eqnarray}
The  first operator relation implies for $q=s$ a relation between $\Delta I = 1$ EWP and 
tree amplitudes in $B\to K\pi$ decays\cite{Neubert:1998pt}.


\begin{thebibliography}{99}
\bibitem{Barberio:2007cr}
  E.~Barberio {\it et al.}  [Heavy Flavor Averaging Group (HFAG)
 Collaboration],
  arXiv:0704.3575 [hep-ex], regularly updated in 
{\tt http://www.slac.stanford.edu/xorg/hfag}.

\bibitem{Cabibbo:1963yz}
  N.~Cabibbo,
  Phys.\ Rev.\ Lett.\  {\bf 10}, 531 (1963).

\bibitem{Kobayashi:1973fv}
  M.~Kobayashi and T.~Maskawa,
  Prog.\ Theor.\ Phys.\  {\bf 49}, 652 (1973).
  
\bibitem{Gronau:1989ia}
M.~Gronau,
Phys.\ Rev.\ Lett.\  {\bf 63}, 1451 (1989).

\bibitem{Boos:2004xp}
  H.~Boos, T.~Mannel and J.~Reuter,
  Phys.\ Rev.\  D {\bf 70}, 036006 (2004)
  [arXiv:hep-ph/0403085];
  M.~Ciuchini, M.~Pierini and L.~Silvestrini,
  Phys.\ Rev.\ Lett.\  {\bf 95}, 221804 (2005)
  [arXiv:hep-ph/0507290];
  H.~n.~Li and S.~Mishima,
  JHEP {\bf 0703}, 009 (2007)
  [arXiv:hep-ph/0610120].

\bibitem{Grinstein:1989df}
  B.~Grinstein,
  Phys.\ Lett.\  B {\bf 229}, 280 (1989).

\bibitem{Ciuchini:1997zp}
  M.~Ciuchini, E.~Franco, G.~Martinelli, A.~Masiero and L.~Silvestrini,
  Phys.\ Rev.\ Lett.\  {\bf 79}, 978 (1997)
  [arXiv:hep-ph/9704274].
 
\bibitem{Xing:1999yx}
  Z.~z.~Xing,
  Phys.\ Rev.\  D {\bf 61}, 014010 (2000)
  [arXiv:hep-ph/9907455]. For an earlier model-dependent estimate see
  Z.~z.~Xing and D.~s.~Du,
  Phys.\ Lett.\  B {\bf 261}, 315 (1991).

\bibitem{Grossman:1996ke}
  Y.~Grossman and M.~P.~Worah,
  Phys.\ Lett.\  B {\bf 395}, 241 (1997)
  [arXiv:hep-ph/9612269].

\bibitem{Aubert:2007pa}
  B.~Aubert {\it et al.}  [BaBar Collaboration],
  Phys.\ Rev.\ Lett.\  {\bf 99}, 071801 (2007)
  [arXiv:0705.1190 [hep-ex]].

\bibitem{Fratina:2007zk}
  S.~Fratina {\it et al.},
  Phys.\ Rev.\ Lett.\  {\bf 98}, 221802 (2007)
  [arXiv:hep-ex/0702031].

\bibitem{Aubert:2006ia}
  B.~Aubert {\it et al.}  [BABAR Collaboration],
  Phys.\ Rev.\  D {\bf 73}, 112004 (2006)
  [arXiv:hep-ex/0604037].

\bibitem{Adachi}
I. Adachi {\it et al.} [Belle Collaboration],
Phys.\ Rev. D {\bf 77}, 091101(R) (2008)
arXiv:0802.2988 [hep-ex].

\bibitem{Aubert:2005jv}
  B.~Aubert {\it et al.}  [BABAR Collaboration],
  Phys.\ Rev.\ D {\bf 72}, 111101 (2005)
  [arXiv:hep-ex/0510051].
  
\bibitem{Zupanc:2007pu}
  A.~Zupanc {\it et al.} [Belle Collaboration],
  Phys.\ Rev.\  D {\bf 75}, 091102 (2007)
  [arXiv:hep-ex/0703040].
  
\bibitem{Sanda:1996pm}
  A.~I.~Sanda and Z.~z.~Xing,
  Phys.\ Rev.\  D {\bf 56}, 341 (1997)
  [arXiv:hep-ph/9702297].

\bibitem{CKMfitter}
CKMfitter Group (J. Charles et al.),
Eur. Phys. J. C41, 1-131 (2005) [hep-ph/0406184];
updated results and plots available at: http://ckmfitter.in2p3.fr.

\bibitem{UTfit}
UTfit group ( M.~Bona {\it et al.}),
AIP Conf.\ Proc.\  {\bf 881}, 210 (2007);
updated results and plots available at: http://www.utfit.org.

\bibitem{Gronau:2007xg}
  M.~Gronau,
  Int.\ J.\ Mod.\ Phys.\  A {\bf 22}, 1953 (2007)
  [arXiv:0704.0076 [hep-ph]].
  
\bibitem{Buchalla:1995vs}
  G.~Buchalla, A.~J.~Buras and M.~E.~Lautenbacher,
  Rev.\ Mod.\ Phys.\  {\bf 68}, 1125 (1996)
  [arXiv:hep-ph/9512380].

\bibitem{Gronau:1994rj}
  M.~Gronau, O.~F.~Hernandez, D.~London and J.~L.~Rosner,
  Phys.\ Rev.\  D {\bf 50}, 4529 (1994)
  [arXiv:hep-ph/9404283];
M.~Gronau, O.~F.~Hernandez, D.~London and J.~L.~Rosner,
  Phys.\ Rev.\  D {\bf 52}, 6374 (1995)
  [arXiv:hep-ph/9504327];
  M.~Gronau, O.~F.~Hernandez, D.~London and J.~L.~Rosner,
  Phys.\ Rev.\  D {\bf 52}, 6356 (1995)
  [arXiv:hep-ph/9504326].
  
\bibitem{Gronau:2000zy}
  M.~Gronau,
  Phys.\ Lett.\  B {\bf 492}, 297 (2000)
  [arXiv:hep-ph/0008292].
 
\bibitem{Fleischer:1999nz}
  R.~Fleischer,
  Eur.\ Phys.\ J.\  C {\bf 10}, 299 (1999)
  [arXiv:hep-ph/9903455];
  R.~Fleischer,
  Eur.\ Phys.\ J.\  C {\bf 51}, 849 (2007)
  [arXiv:0705.4421 [hep-ph]].
  
  \bibitem{Aubert:2006nm}
  B.~Aubert {\it et al.}  [BABAR Collaboration],
  Phys.\ Rev.\  D {\bf 74}, 031103 (2006)
  [arXiv:hep-ex/0605036].
  
 \bibitem{PDGup} W.-M. Yao {\it et al.} [Particle Data Group], 2007 update,
{\tt http://pdglive.lbl.gov/listings1.brl?exp=Y} and 2008 update.
  
 \bibitem{CDFBsDsDs}
T. Aaltonen {\it et al.} [CDF Collaboration], 
Phys.\ Rev.\ Lett.\ {\bf 100}, 021803 (2008).

\bibitem{Blok:1997yj}
  B.~Blok, M.~Gronau and J.~L.~Rosner,
  Phys.\ Rev.\ Lett.\  {\bf 78}, 3999 (1997)
  [arXiv:hep-ph/9701396].

\bibitem{Chiang:2004nm}
  C.~W.~Chiang, M.~Gronau, J.~L.~Rosner and D.~A.~Suprun,
  Phys.\ Rev.\  D {\bf 70}, 034020 (2004)
  [arXiv:hep-ph/0404073].

\bibitem{Gronau:2004ej}
  M.~Gronau and J.~L.~Rosner,
  Phys.\ Lett.\  B {\bf 595}, 339 (2004)
  [arXiv:hep-ph/0405173].

\bibitem{Bjorken:1988kk}
  J.~D.~Bjorken,
  Nucl.\ Phys.\ Proc.\ Suppl.\  {\bf 11}, 325 (1989).

\bibitem{Beneke:2000ry}
  M.~Beneke, G.~Buchalla, M.~Neubert and C.~T.~Sachrajda,
  Nucl.\ Phys.\  B {\bf 591}, 313 (2000)
  [arXiv:hep-ph/0006124].

\bibitem{Bauer:1986bm}
  M.~Bauer, B.~Stech and M.~Wirbel,
  Z.\ Phys.\  C {\bf 34}, 103 (1987);
  M.~Neubert and B.~Stech,
  Adv.\ Ser.\ Direct.\ High Energy Phys.\  {\bf 15}, 294 (1998)
  [arXiv:hep-ph/9705292].

\bibitem{Bortoletto:1990np}
  D.~Bortoletto and S.~Stone,
  Phys.\ Rev.\ Lett.\  {\bf 65}, 2951 (1990).

\bibitem{Rosner:1990xx}
  J.~L.~Rosner,
  Phys.\ Rev.\  D {\bf 42}, 3732 (1990);
  Z.~Luo and J.~L.~Rosner,
  Phys.\ Rev.\  D {\bf 64}, 094001 (2001)
  [arXiv:hep-ph/0101089].
  
\bibitem{Abe:2001yf}
  K.~Abe {\it et al.}  [Belle Collaboration],
  Phys.\ Lett.\  B {\bf 526}, 258 (2002)
  [arXiv:hep-ex/0111082].
  
\bibitem{Rosner:2008yu}
  J.~L.~Rosner and S.~Stone,
  prepared for 2008 Edition of Review of Particle Properties,
  arXiv:0802.1043 [hep-ex].
 
\bibitem{Datta:2003va}
  A.~Datta and D.~London,
  Phys.\ Lett.\  B {\bf 584}, 81 (2004)
  [arXiv:hep-ph/0310252];
  J.~Albert, A.~Datta and D.~London,
  Phys.\ Lett.\  B {\bf 605}, 335 (2005)
  [arXiv:hep-ph/0410015].
  
\bibitem{Gronau:2002qj}
  M.~Gronau and J.~L.~Rosner,
  Phys.\ Rev.\  D {\bf 65}, 093012 (2002)
  [arXiv:hep-ph/0202170].
  
\bibitem{Jenkins:1992qv}
  E.~E.~Jenkins and M.~J.~Savage,
  Phys.\ Lett.\  B {\bf 281}, 331 (1992).
 
\bibitem{Boyd:1995pq}
  C.~G.~Boyd and B.~Grinstein,
  Nucl.\ Phys.\  B {\bf 451}, 177 (1995)
  [arXiv:hep-ph/9502311].
 See also 
  J.~L.~Goity,
  Phys.\ Rev.\  D {\bf 46}, 3929 (1992)
  [arXiv:hep-ph/9206230].
  
\bibitem{Gronau:2007af}
  M.~Gronau and J.~L.~Rosner,
  Phys.\ Lett.\  B {\bf 651}, 166 (2007)
  [arXiv:1704.3459 [hep-ph]].
  
\bibitem{Beneke:2006rb}
  M.~Beneke, M.~Gronau, J.~Rohrer and M.~Spranger,
  Phys.\ Lett.\  B {\bf 638}, 68 (2006)
  [arXiv:hep-ph/0604005].
  
  \bibitem{Gronau:1998fn}
  M.~Gronau, D.~Pirjol and T.~M.~Yan,
  Phys.\ Rev.\  D {\bf 60}, 034021 (1999)
  [Erratum-ibid.\  D {\bf 69}, 119901 (2004)]
  [arXiv:hep-ph/9810482].

\bibitem{Grinstein:1996us}
  B.~Grinstein and R.~F.~Lebed,
  Phys.\ Rev.\  D {\bf 53}, 6344 (1996)
  [arXiv:hep-ph/9602218].

\bibitem{Gronau:2000az}
  M.~Gronau,
  Phys.\ Rev.\  D {\bf 62}, 014031 (2000) 
[arXiv:hep-ph/9911429].

\bibitem{Neubert:1998pt}
  M.~Neubert and J.~L.~Rosner,
  Phys.\ Lett.\  B {\bf 441}, 403 (1998)
  [arXiv:hep-ph/9808493].

\end{thebibliography}
\end{document}